\documentclass[12pt,preprint]{aastex}

\def\ale{\mathrel{\hbox{\rlap{\hbox{\lower4pt\hbox{$\sim$}}}\hbox{$<$}}}}
\def\age{\mathrel{\hbox{\rlap{\hbox{\lower4pt\hbox{$\sim$}}}\hbox{$>$}}}}

\begin{document}

\def\cit{1}
\def\vla{2}

\title{\large A Radio Survey of Type Ib and Ic Supernovae: Searching
for Engine Driven Supernovae}

\author{
E. Berger\altaffilmark{\cit},
S. R. Kulkarni\altaffilmark{\cit},
D. A. Frail\altaffilmark{\vla},
\&\ A. M. Soderberg\altaffilmark{\cit}
}

\altaffiltext{\cit}{Division of Physics, Mathematics and Astronomy,
        105-24, California Institute of Technology, Pasadena, CA
        91125} 
\altaffiltext{\vla}{National Radio Astronomy Observatory, Socorro,
        NM 87801}

\begin{abstract}
The association of $\gamma$-ray bursts (GRBs) and core-collapse
supernovae (SNe) of Type Ib and Ic was motivated by the detection of
SN\,1998bw in the error box of GRB\,980425 and the now-secure
identification of a SN\,1998bw-like event in the cosmological
GRB\,030329.  The bright radio emission from SN\,1998bw indicated that
it possessed some of the unique attributes expected of GRBs, namely a
large reservoir of energy in (mildly) relativistic ejecta and variable
energy input.  The two popular scenarios for the origin of SN\,1998bw
are a typical cosmological burst observed off-axis or a member of a
new distinct class of supernova explosions (gSNe).  In the former,
about 0.5\% of local Type Ib/c SNe are expected to be similar to
SN\,1998bw; for the latter no such constraint exists.  Motivated thus,
we began a systematic program of radio observations of most reported
Type Ib/c SNe accessible to the Very Large Array.  Of the 33 SNe
observed from late 1999 to the end of 2002 at most one is as bright as
SN\,1998bw.  From this we conclude that the incidence of such events
is $\lesssim 3\%$.  Furthermore, analysis of the radio emission
indicates that none of the observed SNe exhibit clear engine
signatures.  Finally, a comparison of the SN radio emission to that of
GRB afterglows indicates that none of the SNe could have resulted from
a typical GRB, independent of the initial jet orientation.  Thus,
while the nature of SN\,1998bw remains an open question, there appears
to be a clear dichotomy between the majority of hydrodynamic and
engine-driven explosions.
\end{abstract}


\section{Introduction}
\label{sec:intro}

The death of massive stars and the processes that lead to the
formation of the compact remnants is a forefront area in stellar
astrophysics.  Recent advances in modeling suggest that great
diversity can be expected.  Indeed, observationally we have already
witnessed a large diversity in the neutron star remnants: radio
pulsars, AXPs, SGRs, and the central source in Cas\,A.  We know
relatively little about the formation of black holes, static or
rotating.

The compact object form following the collapse of the progenitor
core. The energy of the resulting explosion could be supplemented or
even dominated by the energy released from the compact object (e.g.~a
rapidly rotating magnetar or an accreting black hole).  Such
``engines'' can give rise to asymmetrical explosions \citep{mw99}.
Regardless of the source of energy, a fraction of the total energy,
$E_K$, is coupled to the debris or ejecta (mass $M_{\rm ej}$) and it
is these two gross parameters which determine the appearance and
evolution of the resulting explosion.  Equivalently one may consider
$E_K$ and the mean initial speed of ejecta, $v_0$, or the Lorentz
factor, $\Gamma_0=[1-\beta_0^2]^{-1/2}$, where $\beta_0=v_0/c$.

Supernovae (SNe) and $\gamma$-ray bursts (GRBs), are distinguished by
their ejecta velocities.  In the former $v_0\sim 10^4$ km s$^{-1}$ as
inferred from optical absorption features (e.g.~\citealt{fil97}),
while for the latter $\Gamma_0\gtrsim 100$, inferred from the
non-thermal prompt emission \citep{goo86,pac86}, respectively.  The
large difference in initial velocity arises from significantly
different ejecta masses: $M_{\rm ej}\sim few$ M$_\odot$ in SNe
compared to $\sim 10^{-5}$ M$_\odot$ in GRBs.

In the conventional interpretation, $M_{\rm ej}$ for SNe is large
because $E_K$ is primarily derived from the (essentially) symmetrical
collapse of the core and the energy thus couples to all the mass left
after the formation of the compact object.  Mysteriously, $E_K$
clusters around 1 FOE (FOE is $10^{51}$ erg) in most SNe, a mere $1\%$
of the energy released in the gravitational collapse of the core.

Whereas the initial ejecta speed is solely determined by $E_K$ and
$M_{\rm ej}$, a fraction of the ejecta is accelerated to higher
velocities as the blast wave races down the density gradient of the
stellar enveloped (e.g.~\citealt{mm99}).  For the wind- or
binary-stripped (e.g.~\citealt{uom86,bnf91,wlw93,nyp94}) helium and
carbon progenitors of Type Ib and Ic SNe, both factors (a smaller core
mass and a steep density gradient) conspire to produce ejecta at
velocities as high as $\Gamma\beta\sim 1$.  However, only $\lesssim
10^{-5}E_{K}$ is carried by these ejecta \citep{col68,ww86,mm99}.  In
contrast, high velocity ejecta is neither expected nor observed in
Type II SNe with their massive stellar envelopes.

GRB models, on the other hand, appeal to a stellar mass black hole
remnant, which accretes matter on many dynamical timescales and powers
relativistic jets (the so-called collapsar model;
\citealt{woo93,mw99}); highly magnetized neutron stars have also been
proposed (e.g.~\citealt{rtk00}).  Observationally, this model is
supported by the association of some GRBs with SN explosions
(e.g.~\citealt{smg+03}). In addition, the complex temporal profiles
and long duration of GRBs are interpreted in terms of an engine that
is relatively long lived (i.e.~not a singular explosion).  The high
Lorentz factors, a high degree of collimation with opening angles of a
few degrees \citep{fks+01}, and episodes of energy addition presumably
from shells of ejecta with varying Lorentz factors, further
distinguish GRBs from Type Ib/c SNe.

We now recognize that engine-driven events -- GRBs and the recently
discovered X-ray Flashes \citep{hzk03} -- in fact have a wide
dispersion in their ultra-relativistic output as manifested by their
beaming-corrected $\gamma$-ray energies \citep{bfk03} and X-ray
luminosities \citep{bkf03}.  However, these cosmological explosions
appear to have a nearly constant total explosive yield when taking
into account the energy in mildly relativistic ejecta \citep{bkp+03}.

The unusual and nearby ($d\sim 40$ Mpc) SN\,1998bw shares some of the
unique attributes expected of GRBs. This Type Ic SN was found to be
coincident in time and position with GRB\,980425 \citep{gvv+98}, an
event with a single smooth profile.  The inferred isotropic energy in
$\gamma$-rays of GRB\,980425 was only $8\times 10^{47}$ erg
\citep{paa+00}, three to six orders of magnitude fainter than typical
GRBs.  More importantly, SN\,1998bw exhibited unusually bright radio
emission indicating about $10^{50}$ erg of mildly relativistic ejecta
\citep{lc99}.  Equally significant, the radio emission indicated a
clear episode of energy addition \citep{lc99}.  None of these features
-- $\gamma$-rays, significant energy with $\Gamma\beta\gtrsim 2$, and
episodes of energy addition -- have been seen in any other nearby SN.
Thus, the empirical data strongly favor an engine origin for
SN\,1998bw.  In the optical, SN\,1998bw also appears to be extreme:
velocity widths approaching 60,000 km s$^{-1}$ were seen at early time
\citep{imn+98} and the inferred explosion energy may be above normal
values, with estimates ranging from 2 to 50 FOE
\citep{hww99,nmn+01}.

The inference of an engine in SN\,1998bw raises two scenarios for its
origin and relation to GRBs.  GRB\,980425 may have been a typical
burst but viewed well away from the jet axis (hereafter, the off-axis
model), thereby resulting in apparently weak $\gamma$-ray emission
despite the great proximity.  Alternatively, SN\,1998bw represents a
different class of SNe.  If so, collapsars can produce very diverse
explosions.

A powerful discriminant between these two scenarios is the expected
rate of SN\,1998bw-like events.  In the off-axis model, the fraction
of Type Ib/c SNe that are powered by a central engine is linked to the
mean beaming factor of GRBs (e.g.~\citealt{fks+01,tp02}).  Recently,
\citet{fks+01} presented the distribution of jet opening angles for a
sample of 15 GRBs, and found that the mean beaming factor is $\langle
f_b^{-1}\rangle\sim 500$; here $f_b=[1-{\rm cos}(\theta_j)]$ is the
beaming fraction, and $\theta_j$ is the collimation angle.  With an
estimated local GRB rate of $\sim 0.5$ Gpc$^{-3}$ yr$^{-1}$
\citep{sch01} compared to a Type Ib/c SN rate of $\sim 4.8\times
10^{4}$ Gpc$^{-3}$ yr$^{-1}$ \citep{mcp+98,cet99,frp+99}, we expect
that $\sim 0.5\%$ of Type Ib/c SNe will be
similar\footnotemark\footnotetext{We note that this fraction may be
somewhat higher in jet models in which the energy and/or Lorentz
factor decrease away from the jet axis.  The exact fraction depends on
the details of the energy and Lorentz factor distribution (Rossi,
Lazzati, \& Rees 2002; Zhang \& Meszaros 2002)\nocite{rlr02,zm02}.} to
SN\,1998bw.  

On the other hand, if SN\,1998bw is not an off-axis cosmological
burst, then the rate of similar events has to be assessed independent
of the GRB rate.  An upper limit can be obtained by assuming that all
Type Ib/c SNe are engine driven highly asymmetric explosions with
SN\,1998bw having the most favorable geometry.  In this context,
\citet{nor02} has argued that of the $1429$ long-duration
BATSE bursts, about 90 events possess similar high-energy attributes
as that of GRB\,980425. This sub-sample may be concentrated along the
super-galactic plane.  If this sub-sample is accepted as distinct from
the cosmological bursts then $\sim 25\%$ of Type Ib/c SNe within 100
Mpc are expected to be events like SN\,1998bw.

Here, we report a comprehensive program of radio observations of
nearby Type Ib/c SNe.  We began this program in 1999 (motivated by
SN\,1998bw) and observed most reported Type Ib/c SNe with the Very
Large Array.  Our basic hypothesis is that (mildly) relativistic
ejecta are best probed by radio observations, as was demonstrated in
the case of SN\,1998bw.  Furthermore, radio observations of Type Ib/c
SNe allow us to directly compare these objects to the radio afterglows
of cosmological GRBs.  Thus, we can empirically (direct comparison of
radio luminosity distributions) and quantitatively (calorimetry via
radio observations) investigate the link, or lack thereof between Type
Ib/c SNe and cosmological GRBs.  As alluded to above, we did not
investigate Type II SNe since the extended envelopes and dense
circumstellar media of their progenitors are reasonably expected to
mask the activity of a putative engine and thus suppress the presence
of mildly relativistic ejecta to which we are most sensitive.

The organization of the paper is as follows. In \S\ref{sec:obs} we
present the details of the observations.  The results are summarized
in \S\ref{sec:stat}, where we investigate the broad radio properties
(\S\ref{sec:lums}), expansion velocities (\S\ref{sec:vels}), and
energies in high velocity ejecta (\S\ref{sec:energy}).  We further
provide a comparison to the radio afterglows of GRBs in
\S\ref{sec:comp} and draw conclusions in \S\ref{sec:conc}.

\section{Observations}
\label{sec:obs}

Table~\ref{tab:data} summarizes the Very Large Array
(VLA\footnotemark\footnotetext{The VLA is operated by the National
Radio Astronomy Observatory, a facility of the National Science
Foundation operated under cooperative agreement by Associated
Universities, Inc.}) observations of Type Ib/c SNe starting in late
1999 and up to the end of 2002.  We observed a total of 33 SNe out of
51 identified spectroscopically during the same period.  The observed
targets were determined solely by the availability of observing time
and optical selection criteria; we did not employ any additional
selection criteria.

In all observations we used the standard continuum mode with $2\times
50$ MHz bands, centered on 1.43, 4.86, or 8.46 GHz.  We used the
sources 3C\,48 (J0137+331), 3C\,147 (J0542+498), and 3C\,286
(J1331+305) for flux calibration, and calibrator sources within $\sim
5^\circ$ of the SNe to monitor the phase.  The data were reduced and
analyzed using the Astronomical Image Processing System
\citep{fom81}.

\section{Population Statistics}
\label{sec:stat}
 
In this section we investigate the ejecta properties and diversity of
the sample.  Results for individual SNe are given in the Appendix.  In
Figure~\ref{fig:lcs} we plot the radio luminosities and upper limits
for Type Ib/c SNe observed in this survey and in the past (SN\,1983N:
\citealt{spw84}; SN\,1984L: \citealt{psw+86,wsp+86}; SN\,1990B:
\citealt{dsw+93}; SN\,1998bw: \citealt{kfw+98}; SN\,2002ap:
\citealt{bkc02}).  The typical delay between the SN explosion and time
of our observations is about 20 days, with four SNe observed with a
delay of over $100$ days.  In addition, three of the SNe are embedded
in host galaxies with strong radio emission.  For these cases, we
adopt upper limits that correspond to the brightness of the galaxy (at
least ten times the root-mean-square noise of the individual image).
Four of the thirty three SNe have been detected. Thus the detection
rate of our experiment with a typical flux density limit of 0.15\,mJy
($3\sigma$) is about $12\%$.

\subsection{Radio Properties of Type Ib/c SNe}
\label{sec:lums}

Figure~\ref{fig:lcs} provides a succinct summary of the radio light
curves of the Type Ib/c SNe.  Two strong conclusions can be
immediately drawn from this Figure.  First, SNe as bright as
SN\,1998bw are rare.  Second, there is significant dispersion in the
luminosities of Type Ib/c SNe, ranging from $L_{\nu,\rm rad}\approx
10^{29}$ erg s$^{-1}$ Hz$^{-1}$ at the bright end (SN\,1998bw) to that
of SN\,2002ap which is fainter by about four orders of magnitude
\citep{bkc02}.  It is curious that SN\,2002ap also happens to be the
nearest Ib/c SN in our sample (Table~\ref{tab:data}).  Six of the
eight Type Ib/c SNe detected in the radio to date cluster in the range
of about $(3-50)\times 10^{26}$ erg s$^{-1}$ Hz$^{-1}$.  This may be
partly due to a selection effect since the typical detection threshold
is about $4\times 10^{26}(d/50\,{\rm Mpc})^2$ erg s$^{-1}$ Hz$^{-1}$.

We also find that 28 of the 29 non-detections are no brighter than 0.1
times the luminosity of SN\,1998bw.  SN\,2002cg appears potentially
brighter than SN\,1998bw only because it is embedded in a radio bright
host galaxy; we are therefore forced to use a $10\sigma$ limit on its
luminosity (Table~\ref{tab:data}).  Thus, the incidence of bright
events like SN\,1998bw is $\lesssim 3\%$.

As with the radio luminosities, the peak times also exhibit great
variation: at 1.4 GHz the emission from SN\,2002ap peaked at about 7
days, while for SN\,2002cj it peaked at about 65 days.  For SN\,1998bw
the initial peak occurred at 15 days, followed by a second peak at
about 40 days.  Similarly, at 8.5 GHz, SN\,1998bw peaked at 12 and 30
days past explosion, SN\,1983N peaked at about 30 days, and SN\,2002ap
is predicted to have peaked at $\sim 1-2$ days (the first observation
at this frequency was taken about 4 days after the SN explosion).

\subsection{Expansion Velocities}
\label{sec:vels}

If the radio emission arises from a synchrotron spectrum peaking at
the self-absorption frequency, $\nu_a$, then the peak time and peak
luminosity directly measure the mean expansion speed \citep{che98}.
This is simply because the self-absorption frequency is sensitive to
the size of the source, while the luminosity is sensitive to the
swept-up mass.  We use Equation 16 of \citet{che98} to evaluate the
average expansion velocities:
\begin{equation}
v_p\approx 3.1\times 10^4 L_{p,26}^{17/36} t_{p,10}^{-1}
\nu_{p,5}^{-1} \,\,{\rm km\,s^{-1}}.
\label{eqn:vel}
\end{equation}
Here, $L_p=10^{26}L_{p,26}$ erg s$^{-1}$ Hz$^{-1}$ is the peak
luminosity, $t_p=10t_{p,10}$ days is the time of peak emission
relative to the SN explosion, and $\nu_p=5\nu_{p,5}$ GHz is the peak 
frequency.  We infer velocities ranging from $v\sim 10^4$ to $10^5$ km
s$^{-1}$ (Figure~\ref{fig:vel}).  Again, as with the luminosities,
SN\,1998bw with $v\sim c$ is an exception.

We note that if free-free absorption plays a significant role, then
$\nu_p$ is only an upper limit to $\nu_a$, and $L_p$ is a lower limit
to the intrinsic peak luminosity.  In this case, the inferred values
of $v_p$ listed above will in fact be a lower limit to the actual
expansion velocity.  However, this is probably not significant for 
Type Ib/c SNe since their compact progenitors have high escape
velocities and therefore fast winds and low circumburst
densities. Indeed, there is no evidence for free-free absorption
either for SN\,1998bw \citep{kfw+98,lc99} or SN 2002ap
\citep{bkc02}.

Estimating the expansion velocities for the non-detections is not 
straightforward since we cannot ensure that the limits constrain the 
peak luminosity.  We are therefore forced to make an additional
assumption.  For example, if we assume that most Type Ib/c SNe are
similar in their emission properties to SN\,1983N, then the majority
of upper limits approximately sample the peak emission and the
inferred upper limits are $\lesssim 0.3c$ (Figure~\ref{fig:vel}).  

On the other hand, if the typical peak time is only a few days then 
our observations only constrain the decaying portion of the lightcurve
and the expansion velocity may be higher.  Fortunately, this is 
not a significant problem based on the following argument.  The 
equipartition energy directly depends on the peak luminosity, $U_{\rm
eq}\approx 3.7\times 10^{46}L_{p,27}^{20/17}$ erg (see
\S\ref{sec:2002cj}), where $L_p=10^{27}L_{p,27}$ erg s$^{-1}$ is the
peak luminosity at 8.5 GHz.  With a typical fading rate of $F_\nu
\propto t^{-1}$ in the optically thin regime, there are a few SNe that
could have reached a peak luminosity of the order of $10^{29}$ erg
s$^{-1}$ if $t_p=1$ day post explosion.  This is a reasonable minimum
peak time taking into account the deceleration time of the ejecta.
Thus, the equipartition energy is at most $10^{49}$ erg, about two
orders of magnitude lower than typical GRBs (\S\ref{sec:conc}).  For
most non-detections the limit is in fact much lower, $\sim 5\times
10^{46}$ to $10^{48}$ erg.  This indicates that a few of the
non-detected sources may have in fact produced mildly relativistic
ejecta, but these would still be energetically uninteresting when
compared to SN\,1998bw let alone GRB afterglows.

\subsection{Energetics}
\label{sec:energy}

In the previous section we found that no SN observed to date is
comparable to SN\,1998bw especially in regard to the mean expansion
speed.  SN\,1998bw was also interesting because it possessed an
unusually large amount of energy in mildly relativistic ejecta.
However, a purely hydrodynamic explosions can also produce {\it some}
amount of relativistic ejecta.  The energy of such ejecta can be
estimated using well understood models of shock propagation in the
pre-supernova cores \citep{che82,mm99}.  The key parameters are $E_K$
and $M_{\rm ej}$ which can be inferred from the optical lightcurves
and spectra using hydrodynamic models of a SN explosion in a CO core
coupled with radiative transfer calculations (e.g.~\citealt{inn+00}).

In this section, we investigate whether any of the detected Type Ib/c
SNe possess such large energy in high velocity ejecta that cannot be
explained by the simplest hypothesis of a purely hydrodynamic
explosion.  To this end, in Table~\ref{tab:ejecta} we summarize the
results of hydrodynamic models for the SNe that have been detected in
the radio.

The ejecta produced in a hydrodynamic explosion has a density profile 
that can be described by power laws at low and high velocities,
separated by a break velocity, which for Type Ib/c progenitors is
given by \citep{mm99}:
\begin{equation}
v_{\rm ej,b}\approx 5.1\times 10^3(E_{K,51}/M_{\rm ej,1})^{1/2}
\,{\rm km\,s^{-1}}. 
\label{eqn:vb}
\end{equation}
Here $E_K=10^{51}E_{K,51}$ erg and $M_{\rm ej}=10M_{\rm ej,1}$
M$_\odot$.  For typical values of $E_K$ and $M_{\rm ej}$, the radio
emission from the detected SNe is produced by ejecta above the break 
velocity.  In particular, for SN\,2002ap, $v_{\rm ej,b}\approx 2\times
10^4$ km s$^{-1}$, which is lower than the velocity of the ejecta
producing the radio emission, $v\approx 9\times 10^4$ km s$^{-1}$
\citep{bkc02}.  Similarly, for SN\,1998bw $v_{\rm ej,b}$ ranges from
about $1.5\times 10^4$ to $3.5\times 10^4$ km s$^{-1}$ (depending on
which model is assumed, Table~\ref{tab:ejecta}) while the radio
emission was produced by ejecta expanding with $\Gamma\beta\approx 2$.

The ejecta velocity profile extends up to a cutoff determined by
significant radiative losses when the shock front breaks out of the
star.  For a radiative stellar envelope this is $v_{\rm ej,max}\approx
11.5\times 10^4 E_{K,51}^{0.58} M_{\rm ej,1}^{-0.42}$ \citep{mm99},
assuming a stellar radius of 1 R$_\odot$.  For the SNe considered
here we find cutoff velocities of $\Gamma\beta\sim 1-3$.  

To determine whether there is sufficient energy in fast ejecta to
account for the radio observations we calculate the energy above a
velocity, $V$ \citep{mm99}:
\begin{equation}
E(v>V)\approx \int_V^\infty \frac{1}{2}\rho_fv^24\pi v^2t^3dv \\
\approx 7.2\times 10^{44}E_{K,51}^{3.59}M_{\rm
ej,1}^{-2.59}V_5^{-5.18} \,{\rm erg},
\label{eqn:energy}
\end{equation}
where $V_5$ is the velocity in units of $10^5$ km s$^{-1}$.  

Unfortunately, as can be seen from Table~\ref{tab:ejecta}, only four
(including SN\,1998bw) SNe have sufficient optical data which is
necessary to estimate $E_K$ and $M_{\rm ej}$.  Of this limited sample,
much of the radio data for SN\,1994I remain unpublished.  Thus we are
left with SN\,2002ap, SN\,1983N, and SN\,1998bw.

Using the parameters given in Table~\ref{tab:ejecta} for SN\,2002ap,
\citet{bkc02} find $E(v>0.3c)\approx 3.8\times 10^{48}$ erg.  In
contrast, from the radio observations we estimate $2\times 10^{46}$
erg.  Thus, there is no need, nor indeed room, for mildly relativistic
ejecta in this SN.  We therefore disagree with the claims of high
velocity jets carrying a large amount of energy, $\sim 0.1-1$ FOE,
made by \citet{kji+02} and \citet{tot03}.  Furthermore, the large
discrepancy between the amount of energy inferred from the
hydrodynamic models and the radio observations suggests that either
the optically-derived parameters are in error,
Equation~\ref{eqn:energy} has an incorrect pre-factor, or the radio
estimate is incorrect.  However, the radio estimate is relatively
robust (eventually related to equipartition energy estimates).  On the
other hand, as with SN\,1998be \citep{hww99} the total kinetic energy
may have been over-estimated, possibly as a result of neglecting a
mild asymmetry.

For SN\,1984L we do not have a direct estimate of the energy in the
radio-emitting ejecta since the peak of the radio emission has been
missed.  However, based on the similarity to SN\,1983N in the
optically thin regime we estimate $L_p(t=30\,{\rm d},\nu_p=5\,{\rm
GHz})\approx 1.4\times 10^{27}$ erg s$^{-1}$ Hz$^{-1}$.  This
translates to a peak flux of $3.2$ mJy at the distance of SN\,1984L
($d\approx 19$ Mpc).  Using the equipartition analysis presented in
\S\ref{sec:2002cj} we estimate an energy of about $7\times 10^{46}$
erg, and an average expansion velocity of about $0.1c$.  From
Equation~\ref{eqn:energy} we find $E(v>0.1c) \approx 3\times 10^{50}$
erg -- similar to the conundrum discussed above for SN\,2002ap.  

For SN\,1998bw, on the other hand, we find $E(v>c)\approx 2\times
10^{45}$ erg using the parameters inferred by \citet{hww99}, or
$E(v>c)\approx 3\times 10^{48}$ erg using the parameters given by
\citet{imn+98}.  In both cases, the energy available in fast ejecta is
significantly lower than the energy inferred from the radio emission,
$\sim 10^{50}$ erg.

To conclude, for SN\,1984L and SN\,2002ap a hydrodynamic explosion can
supply the energy and velocity that are responsible for the observed
radio emission.  Most likely, the same is true for the non-detections.
On the other hand, SN\,1998bw is a clear exception, exhibiting a
significant excess of energy in ejecta moving with $\Gamma\beta\approx
2$ compared to what is available from a hydrodynamic explosion.

\section{A Comparison to $\gamma$-Ray Burst Afterglows}
\label{sec:comp}

\begin{figure} 
\plotone{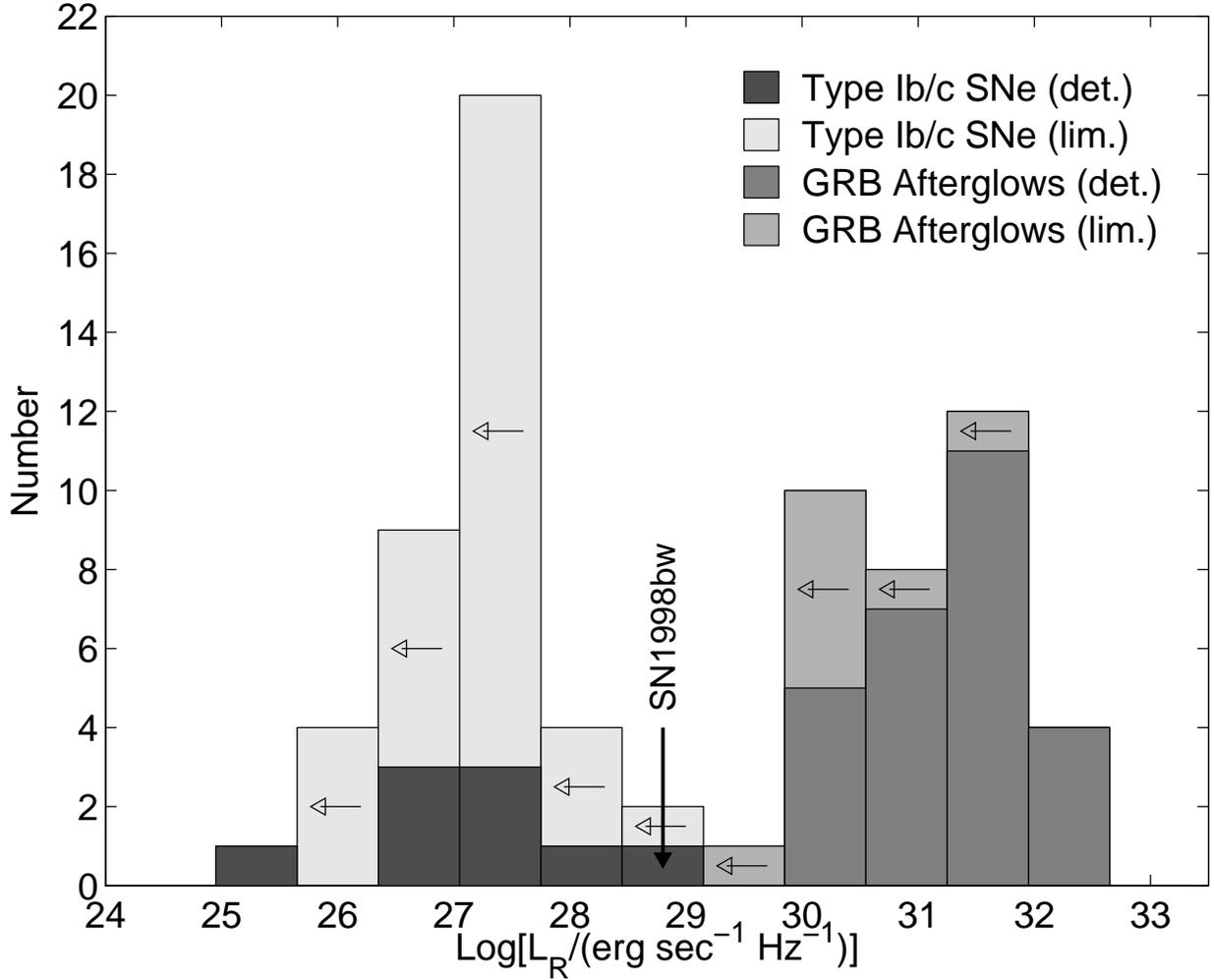}
\caption{Histograms of the radio luminosity of Type Ib/c SNe from
this survey and the literature, and GRB radio afterglows from the
sample of Frail et al. (2003).  Upper limits are plotted as $3\sigma$,
unless there is significant contamination from the host galaxy (see
Table~\ref{tab:data}).  
\label{fig:hist1}}
\end{figure}

In the previous section we investigated the radio properties of Type
Ib/c SNe and found that in every respect SN\,1998bw was unique.  In
this section we compare the Ib/c sample (including SN\,1998bw) with
the radio afterglows of GRBs.  In Figure~\ref{fig:lcs2} we plot the
radio lightcurves of GRB\,970508 \citep{fwk00}, a typical cosmological
burst, and the nearest event, GRB 030329 \citep{bkp+03}, in addition
to the SN lightcurves.  As demonstrated by this figure and
Figure~\ref{fig:hist1}, the radio lightcurves of GRB afterglows and
SNe are dramatically different.  Furthermore, SN\, 1998bw is unique in
both samples: it is fainter than typical radio afterglows of GRBs but
much brighter than Type Ib/c SNe.

Figures~\ref{fig:lcs2} and \ref{fig:hist1} have significant
implications, namely {\it none} of the Type Ib/c SNe presented here
could have given rise to a typical $\gamma$-ray burst. It has been
suggested that GRBs are distant Type Ib/c SNe but with their jets
pointed at the observer, whereas such a bias is absent in the nearby
Type Ib/c sample. However, most of our radio observations are obtained
on a timescale of 10--100 days (see Figure~\ref{fig:lcs}).  Scaling
from the observed ``jet'' break times of a few days in GRB afterglows,
off-axis collimated explosions become spherical on a timescale of
$\sim 10-10^2$ days \citep{pac01,gl03} at which point the relative
geometry between the observer and the explosion is not important.
Thus, we find no evidence suggesting that all or even a reasonable
majority of Type Ib/c SNe give rise to GRBs.  We now quantify the
difference between the Ib/c and GRB samples.

Our goal here is to determine the differential luminosity
distribution, $n(L)$, which agrees with both detections and upper
limits; here $n(L)$ is the number of events with luminosity between
$L$ and $L+dL$.  It is important to include non-detections since they
represent the majority of the SN data.  Similarly, we include upper
limits on the radio luminosity of GRB afterglows that have been
localized in other wave-bands (i.e.~optical and X-rays) and for which
a redshift has been measured.  Unfortunately, as many as half of the
GRBs localized in the X-rays do not have a precise position, and hence
a redshift.  For these afterglows it is not possible to provide a
limit on the radio luminosity.  Still, with a typical flux limit of
about 0.3 mJy ($5\sigma$; \citealt{fkb+03}), and assuming that these
sources have a similar distribution of redshifts to the detected
afterglows, we find typical luminosity limits of about $10^{31}$ erg
s$^{-1}$ Hz$^{-1}$, consistent with the peak of the distribution of
detected afterglows. Therefore, unless these sources are biased to low
redshift we do not expect a strong bias as a result of neglecting
them.

The quality of fit for $n(L)$  is determined using the Likelihood
function, ${\cal L}=\prod_{i=1}^{N} {\cal L}_i$, with \citep{ry01}: 
\begin{equation}
{\cal L}_i=\cases{\int_{-\infty}^{\infty}n(L){\cal G}
(L_i,\sigma_{Li})dL & $L_i=$ detection \cr
\int_{L_i}^{\infty}n(L)dL & $L_i=$ limit},
\end{equation}  
where $N$ is the total number of sources (SNe or afterglows), and
${\cal G}(L_i,\sigma_{Li})$ is a normalized Gaussian profile
centered on the observed luminosity of a detected source and with a 
width equal to the $1\sigma$ rms uncertainty in the luminosity.  

\begin{figure} 
\epsscale{0.8}
\plotone{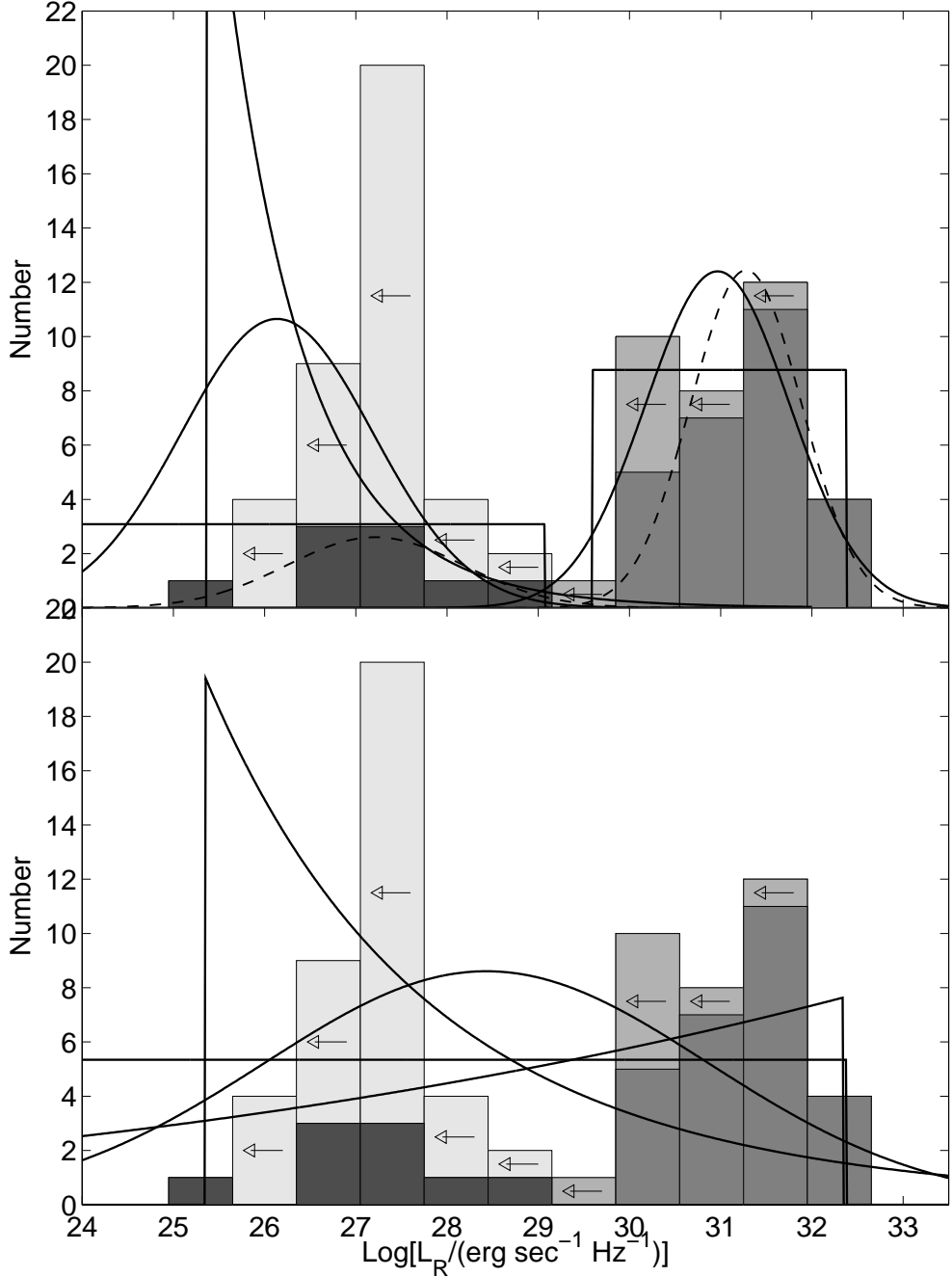}
\caption{Same as Figure~\ref{fig:hist1} but with models of the
luminosity distribution (\S\ref{sec:comp}).  The dashed lines are a
Gaussian profile fit for the detections in each sample.  The solid
lines are a fit to the detections and upper limits using several
models for the distribution function (\S\ref{sec:comp}).  In the top
panel we model each population separately, whereas the bottom panel
shows models for the combined populations.  No single distribution can
fit both the local Type Ib/c SN population and the cosmological GRB
population (Table~\ref{tab:bayes}).
\label{fig:hist2}}
\end{figure}

We consider four models for $n(L)$ based on the apparent distribution
of the detections and upper limits: a Gaussian,
\begin{equation}
n(L)=\frac{1}{\sqrt{2\pi\sigma_L}}\,{\rm
exp}{\left[-\frac{1}{2}\left(\frac{L-L_0}{\sigma_L}\right)^2\right]}, 
\label{eqn:gauss}
\end{equation}
a decreasing power-law,
\begin{equation}
n(L)=\cases{0 & $L<L_0$ \cr
(1-\alpha_L)L^{\alpha_L}/(L_0^{\alpha_L+1}) & $L\ge L_0$},
\label{eqn:pl}
\end{equation}
an increasing power-law,
\begin{equation}
n(L)=\cases{(1+\alpha_L)L^{\alpha_L}/(L_0^{\alpha_L+1}) & $0\le L<L_0$
\cr 
0 & $L\ge L_0$},
\label{eqn:ipl}
\end{equation}
and a flat distribution,
\begin{equation}
n(L)=\cases{0 & $L<L_1$ \cr 1/(L_2-L_1) & $L_1\le L\le L_2$ \cr 0 &
$L>L_2$}.
\label{eqn:box}
\end{equation}
In each case we fit for the two free parameters (e.g.~$L_0$ and
$\sigma_L$ in Equation~\ref{eqn:gauss}).  We do not use the increasing
power law model for the individual distributions since the
observations are clearly inconsistent with such a model.  The
resulting best-fit models are shown in Figure~\ref{fig:hist2} and
summarized in Table~\ref{tab:bayes}.   

We find that the SN population is modeled equally well with the
Gaussian, flat, or decreasing power law distributions, while the GRB
afterglows can be fit with a Gaussian or flat distributions; a
decreasing power law provides a much poorer fit.  Regardless of the
exact distribution the two populations require distinctly different
parameters, with a minimal overlap at the tails of the distributions.

Fitting the combined SN and afterglow data with the models provided
above (Figure~\ref{fig:hist2} and Table~\ref{tab:bayes}) we find that
even the best models (an increasing power-law or a flat distribution)
provide a much poorer fit; the likelihood of the fits are ${\rm
ln}({\cal L})\approx 104$ compared to the combined value of $61.3$ for
the separate Gaussian fits.  Thus, the two populations cannot be
accommodated with a simple single distribution.  This points to a
separate origin for the GRB and Type Ib/c SN populations.  However,
SN\,1998bw can be accommodated in either population. It is equally
plausibe that it is a low luminosity GRB or the brightest Type Ib/c
radio supernova known to date.

\section{Discussion and Conclusions}
\label{sec:conc}

We presented VLA radio observations of 33 Type Ib/c SNe observed
between late 1999 and the end of 2002.  Four of these SNe have been
detected, giving a detection rate of about $12\%$ above a typical
$3\sigma$ flux limit of 0.15 mJy.  At the same time, the combined
detections and non-detections indicate that at most $3\%$ of Type Ib/c
SNe are as luminous as SN\,1998bw, although the single source which
may be brighter is only so because it is embedded in a radio bright
host galaxy.

We infer typical velocities of the radio-emitting ejecta of about
$10^4-10^5$ km s$^{-1}$ for the detected SNe, and upper limits in the
same range for the non-detections.  We also find that a hydrodynamic
explosion can supply the energy carried by the fastest ejecta.
Finally, none of the detected SNe show clear evidence for variable
energy input (shells with different velocity or continued activity by
the central engine); however, we note that our sampling is quite
sparse.

The measurements (radio lightcurves) and inferences (energy in fast
ejecta, energy addition) offer no compelling reason to conclude that
any of our SNe have the special properties of SN\,1998bw
(\S\ref{sec:bw}).

\citet{nor02} has proposed, based on the empirical
lag-luminosity relation, that 25\% of Type Ib/c SNe are similar to
SN\,1998bw.

We also compared the Type Ib/c sample with the sample of radio
afterglow of GRBs. Empirically, these two populations appear to be
quite disparate.  This conclusion is reinforced by careful modeling of
the luminosity distributions.  Still, SN\,1998bw may belong to either
population.

\subsection{What is SN\,1998bw?} 
\label{sec:bw}

Our four year survey of Type Ib/c SNe was first and foremost motivated
by the peculiar object, SN\,1998bw. This supernova showed three
attributes unique to GRBs: relativistic ejecta, substantial reservoir
of energy in such ejecta, and energy addition.  A singular
hydrodynamic explosion cannot account for these attributes.  A natural
explanation is that SN\,1998bw, like GRBs, was ,driven by an engine
powerful enough to significantly modify the explosion.

Our survey has demonstrated that SN\,1998bw-like events are rare in
the local sample.  This begs the question: what is SN\,1998bw?

Two popular scenarios have been suggested.  The first -- the
``off-axis'' scenario -- holds that SN\,1998bw is a typical GRB albeit
nearby and with collimated ejecta pointed away from us
\citep{hww99,mw99,nak99,gpk+02}.  This hypothesis is attractive
because of its simplicity.  We know GRBs exist and most of them do not
point towards us \citep{fks+01}.

\begin{figure} 
\epsscale{0.8}
\plotone{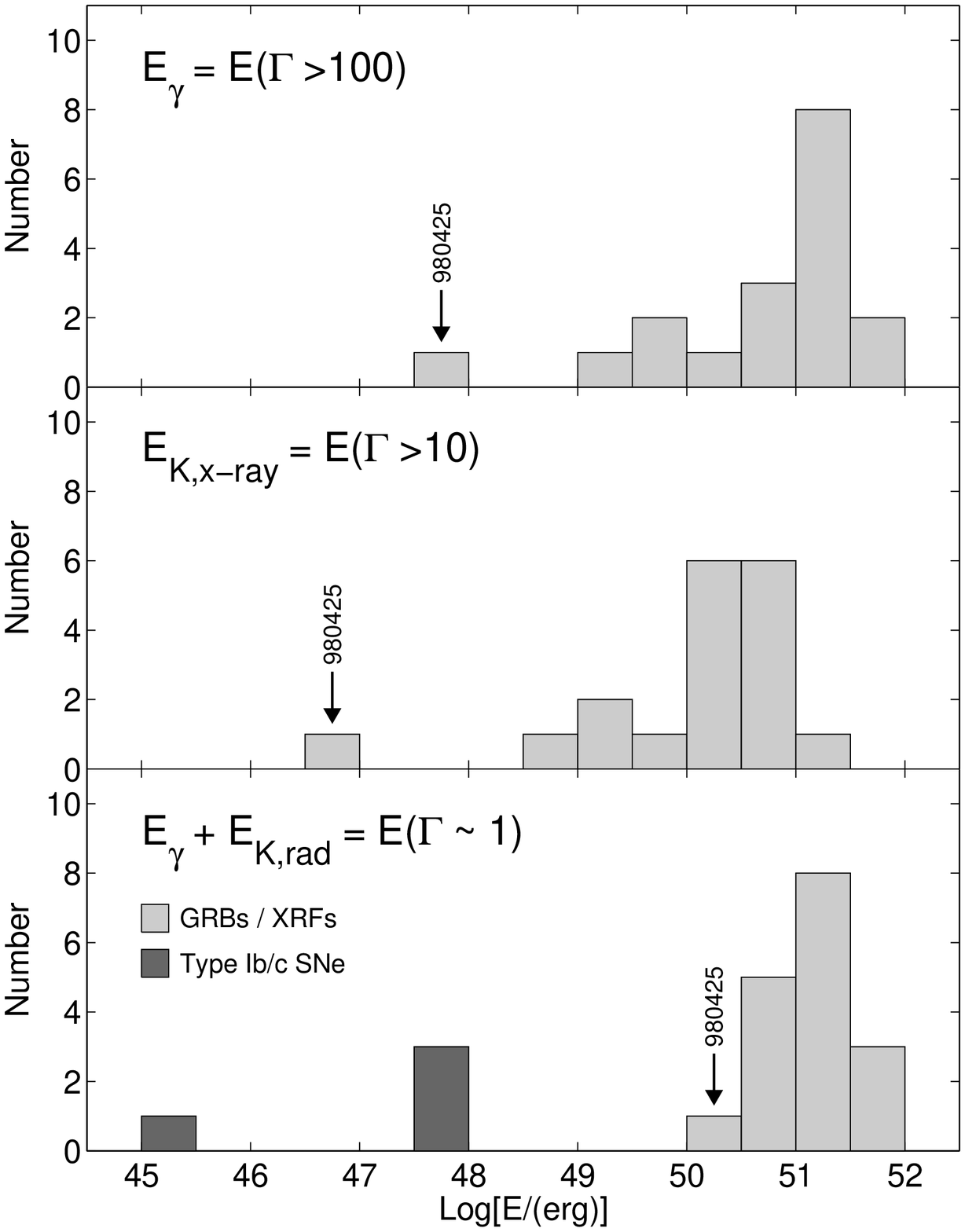}
\caption{Histograms of the beaming-corrected $\gamma$-ray energy
\citep{bfk03}, $E_\gamma$, the kinetic energy inferred from X-rays at
$t=10$ hr \citep{bkf03}, $E_{K,X}$, and total relativistic energy,
$E_\gamma+E_{K}$, where $E_K$ is the beaming-corrected kinetic energy
inferred from the broad-band afterglows of GRBs \citep{lc99,pk02} and
radio observations of SNe.  The wider dispersion in $E_\gamma$ and
$E_{K,X}$ compared to the total energy indicates that engines in
cosmic explosions produce approximately the same quantity of energy,
thus pointing to a common origin, but the ultra-relativistic output of
these engines varies widely.  In Type Ib/c SNe, on the other hand, the
total explosive yield in fast ejecta (typically $\sim 0.3c$) is
significantly lower.  This points to a separate origin for these two
explosive phenomena.
\label{fig:hist3}}
\end{figure}

In the other scenario SN\,1998bw is a new type of explosion with
little energy in ultra-relativistic ejecta \citep{bkh+98,kfw+98}.
Evidence in favor of this idea is best illustrated by
Figure~\ref{fig:hist3} where we find that GRB\,980425 is consistently
at the faint end of the GRB population.

Unfortunately, we are not able to decisively resolve this controversy.
As demonstrated by Figure~\ref{fig:hist1}, one could argue that
SN\,1998bw is at the bright end of the radio luminosity function of
Type Ib/c supernovae or at the faint end of GRB radio afterglow.

The expected rate of SN\,1998bw-like events in the off-axis framework
is about $0.5\%$ of Type Ib/c events, given the average beaming factor
of about $500$ derived by \citet{fks+01}.  Thus, it is not entirely
improbable that one out of about 40 Type Ib/c SNe observed to date is
an off-axis GRB.  As an aside, we can use our $3\%$ limit and compare
the event rate of Type Ib/c SNe with the observed rate of GRBs
(\S\ref{sec:intro}) to place a limit of $f_b\gtrsim 3\times 10^{-4}$
on the beaming fraction.  This corresponds to a limit of
$\theta_j\gtrsim 1.4^\circ$ on the jet opening angles of GRBs;
narrower jets are not likely.  This result may also be interpreted as
a limit on angular size of the highly relativistic core in models of
variable energy and/or Lorentz factor across the surface of the jet
\citep{rlr02,zm02}.

In the gSN framework, we now know that at most a few percent of Type
Ib/c are possibly gSNe.  At the same time, the recent GRB\,030329 was
accompanied by a SN similar to SN\,1998bw (SN\,2003dh;
\citealt{smg+03,hjo+03}).  Thus, an investigation of the number and
properties of gSNe requires observations of both local Type Ib/c SNe
and GRBs.

While we cannot determine the exact origin of SN\,1998bw based on the 
statistics of our survey, the ultimate detection of similar events at
the level of about $1\%$ may in fact allow us to distinguish between
the off-axis and gSN scenarios.

\subsection{Hypernovae}

The discovery of broad optical lines in SN\,1998bw and large explosive
energy release, $\gtrsim {\rm few}$ FOE, prompted some astronomers to
use the designation ``hypernovae'' for SN\,1998bw-like SNe.
Unfortunately, this designation is not well defined.  To begin with,
the term hypernova was first used by \citet{pac98} to describe the
GRB/afterglow phenomenon; thus, this term implies a connection to
GRBs.  The prevalent view now is that hypernovae are characterized by
broad optical absorption lines and larger than normal energy release.
However, neither of these criteria has been defined quantitatively by
their proponents.

Ignoring this important issue, the following have been suggested to be
hypernovae: the Type Ib/c SNe 1992ar \citep{cps+00}, 1997dq
\citep{mfl+01}, 1997ef \citep{inn+00,min00}, 1998ey \citep{gjk+98}, and
2002ap \citep{mdm+02}, and the Type II SNe 1992am \citep{ham03}, 1997cy
(e.g.~\citealt{grs+00}), and 1999E \citep{rtb+03}.  Some have also been
claimed to be associated with GRBs detected by BATSE, but at a low
significance.

Our view is that the critical distinction between an ordinary
supernova and a GRB explosion are relativistic ejecta carrying a
considerable amount of energy.  Such ejecta are simply not traced by
optical spectroscopy.  This reasoning is best supported by the fact
that the energy carried by the fast ejecta in SN\,1998bw and
SN\,2002ap \citep{bkc02} differ by four orders of magnitude even
though both exhibit broad spectral features at early times.  Thus,
broad lines do not appear to be a good surrogate for SN\,1998bw-like
objects.

In addition, in two cases, SNe 2002ap and 1984L, the energy inferred
from the radio observations indicates that the total kinetic energy as
inferred by optical spectroscopy and light curves may have been
over-estimated by an order of magnitude (\S\ref{sec:energy}).  It is
possible that "hypernovae" are in fact only slightly more energetic
than typical Type Ib/c SNe, but exhibit a mild degree of asymmetry,
leading to excessively high estimates of the total energy when a
spherical explosion is assumed.

We suggest that the term hypernova be reserved for those SNe, like
SN\,1998bw, which show direct evidence for an engine through the
presence of relativistic ejecta.  As illustrated by SN\,1998bw, the
relativistic ejecta are reliably traced by radio observations.

We end with the following conclusions.  First, radio observations
provide a robust way of measuring the quantity of energy associated
with high velocity ejecta.  This allows us to clearly discriminate
between engine-driven SNe such as SN\,1998bw and ordinary SNe, powered
by a hydrodynamic explosion, such as SN\,2002ap \citep{bkc02} and SNe
2001B, 2001ci, and 2002cj presented here. Second, since at least
$97\%$ of local Type Ib/c SNe are not powered by inner engines and
furthermore have a total explosive yield of only $10^{48}$ erg in fast
ejecta (where fast means $v\sim 0.3c$ compared to $\Gamma\sim few$ in
GRBs), there is a clear dichotomy between Type Ib/c SNe and cosmic,
engine-driven explosions (Figure~\ref{fig:hist3}).  The existence of
intermediate classes of explosions and the nature of SN\,1998bw may be
ascertained with continued monitoring of several hundred Type Ib/c SNe
and cosmological explosions.  Fortunately, such samples will likely
become available over the next few years.

\acknowledgements
We thank B.~Schmidt for helpful comments.  GRB and SN research at
Caltech is supported in part by funds from NSF and NASA.

\appendix

\section{Results for Individual Supernovae}
\label{sec:ind}

\subsection{SN\,2001B}
\label{sec:2001B}

SN\,2001B was discovered in images taken on 2001, Jan 3.61 and 4.57
UT, approximately 5.6 arcsec west and 8.9 arcsec south of the nucleus
of IC\,391 \citep{xq01-iauc7555}.  The SN explosion occurred between
2000, Dec 24.54 UT and the epoch of discovery.  Based on an initial
spectrum, taken on 2001, Jan 14.18 UT \citet{mjc+01-iauc7563}
concluded that the SN was of Type Ia approximately $9$ days past
maximum brightness, showing well defined Si\,II and Ca\,II features,
with a Si\,II expansion velocity of about 7400 km sec$^{-1}$.  A
subsequent spectrum taken on 2001, Jan 23 indicated that the SN was in
fact of Type Ib based on clear He\,I absorption lines
\citep{cf01-iauc7577}.    

We observed SN\,2001B on 2001, Feb 4.49 UT at 8.46 GHz.  An initial
analysis revealed a source which was interpreted as the host galaxy of
the SN, with a flux of $3.5$ mJy.  A second epoch obtained on 2002,
Oct 28.45 UT revealed that the source has faded below $0.12$ mJy,
establishing it as the radio counterpart of SN\,2001B.
(Figure~\ref{fig:lcs}).

\subsection{SN\,2001ci}
\label{sec:2001ci}

SN\,2001ci was initially detected in images taken with the Katzman
Automatic Imaging Telescope on 2001 Apr 25.2 UT, 6.3 arcsec west and
25.4 arcsec north of NGC\,3079 \citep{slf01-iauc7618}.  These 
observations did not provide conclusive evidence that the source was
in fact a SN.  A spectrum obtained by \citet{fc01-iauc7638} on 2001,
May 30 UT revealed that the source was in fact a Type Ic SN about
$2-3$ weeks past maximum brightness \citep{mfl+01}.  

We observed the SN on 2001, Jun 6.35 UT at 8.46 GHz, but did not
detect the source since it appeared to be part of the host galaxy
extended structure.  A second epoch taken on 2002, Jun 10.9 UT
revealed a clear fading at the optical position of the SN with a flux
of $1.45\pm 0.25$ mJy in the first epoch, and a $2\sigma$ limit of
$0.3$ mJy in the second epoch.

\subsection{SN\,2002cj}
\label{sec:2002cj}

SN\,2002cj was discovered on 2002, Apr 21.5 UT, 1.4 arcsec west and
3.9 arcsec south of the nucleus of ESO 582-G5, at a distance of 106
Mpc \citep{gl02-iauc7882}.  The SN explosion occurred between Apr 9.5
UT and the epoch of discovery.  Spectra of the SN obtained on 2002,
May 2.43 and May 7 UT revealed that SN\,2002cj is of Type Ic
\citep{mjc+02-iauc7894,cho02-iauc7894}.

We initially observed the SN on 2002, Jun 3.19 UT at 1.43, 4.86, and
8.46 GHz and detected a faint source at all three frequencies.  The
SN spectrum is given by $F_\nu\propto\nu^{-0.1\pm 0.3}$ between 1.43
and 4.86 GHz, and $F_\nu\propto\nu^{-1.2\pm 0.8}$ between 4.86 and
8.46 GHz, indicating that the spectral peak was most likely located
between 1.43 and 4.86 GHz during the first epoch ($\Delta t\approx
43-55$ days).  In subsequent observations at 1.43 GHz the SN
brightened and then faded, as expected if the peak was in fact above
1.43 GHz initially, and shifted through the band at later epochs.
Using the expected shape of the spectrum, $F_\nu\propto\nu^{5/2}$ for
$\nu<\nu_p$ and $F_\nu \propto\nu^{-(p-1)/2}$ for $\nu>\nu_p$ (with
$p\sim 3$), we find $F_{\nu,p}\sim 0.5$ mJy at $\nu_p\sim 2$ GHz and
$\Delta t=43-55$ days.   

We use these values along with the well-established equipartition
analysis \citep{rea94} to derive some general constraints on the
properties of the
emitting material.  In particular, the energy of a synchrotron source
with flux density, $F_p(\nu_p,t_p)$, can be expressed in terms of the
equipartition energy density,   
\begin{equation}
\frac{U}{U_{\rm eq}}=\frac{1}{2}\epsilon_B\eta^{11} 
(1+\frac{\epsilon_e}{\epsilon_B}\eta^{-17}),
\end{equation}
where $\eta=\theta_s/\theta_{\rm eq}$, $\theta_{\rm eq}\approx
120d_{\rm Mpc}^{-1/17}\,F_{p,\rm mJy}^{8/17}\,\nu_{p, \rm GHz}^{(-2
\beta-35)/34}$ $\mu$as, $U_{eq}=1.1\times 10^{56}d_{\rm
Mpc}^{2}\,F_{p,\rm mJy}^{4}\,\nu_{p,\rm GHz}^{-7}\,\theta_{\rm eq,\mu
as}^{-6}$ erg, and $\epsilon_e$ and $\epsilon_B$ are the fractions of
energy in the electrons and magnetic fields, respectively.  In
equipartition $\epsilon_e=\epsilon_B=1$ and the energy is minimized; a
deviation from equipartition will increase the energy significantly.

For SN\,2002cj we find $\theta_{\rm eq}(t=43-55\,{\rm d})\approx 30$
$\mu$as (i.e.~$R_{\rm eq}\approx 5\times 10^{16}$ cm), which indicates
an average expansion velocity, $v_{\rm eq}\approx (0.35-0.45)c$.  The
equipartition energy is $U_{\rm eq}\approx 8\times 10^{47}$ erg,
indicating a magnetic field strength of $B_{\rm eq}\approx 0.2$ G.

\clearpage

\begin{deluxetable}{llllllcccc}
\rotate
\tablecolumns{10}
\tablewidth{0pc}
\tabletypesize{\footnotesize}
\tablecaption{Radio Observations of Type Ib/c Supernovae in the Period
1999-2002 \label{tab:data}} 
\tablehead{
\colhead {SN} &
\colhead {IAUC} &
\colhead {$t_0$} &
\colhead {$t_{\rm obs}$} &
\colhead {$\Delta t$} &
\colhead {Dist.} &
\colhead {Detected?} &
%\colhead {$F_{22.5}$} &
\colhead {$F_{8.46}$} &
\colhead {$F_{4.86}$} &
\colhead {$F_{1.43}$} \\
\colhead {} &
\colhead {} &
\colhead {(UT)} &
\colhead {(UT)} &
\colhead {(days)} &
\colhead {(Mpc)} &
\colhead {} &
%\colhead {($\mu$Jy)} &
\colhead {($\mu$Jy)} &
\colhead {($\mu$Jy)} &
\colhead {($\mu$Jy)} 
}
\startdata
%1999P  & 7097 &  &  &  &  &  \\
%1999bc & 7133 &  &  &  &  &  \\
%1999bu & 7145 &  &  &  &  &  \\
%1999bv & 7148 &  &  &  &  &  \\ 
%1999bz & 7157 &  &  &  &  &  \\ 
%1999cn & 7202 &  &  &  &  &  \\ 
%1999di & 7234 &  &  &  &  &  \\ 
%1999dn & 7241 &  &  &  &  &  \\ 
%1999ec & 7268 &  &  &  &  &  \\ 
%1999eh & 7282 &  &  &  &  &  \\ 
1999ex & 7310 & Oct 25.6--Nov 9.5 & Nov 18.02 & 8.5--23.4 & 54 & No
& $\pm 53^{a}$ & $\pm 71$ & \nodata \\ \\
2000C  & 7348 & 1999 Dec 30--2000 Jan 8.3 & Feb 4.04 & 26.7--36.0 & 59
& Hint & \nodata & $187\pm 84$ & \nodata \\ 
2000F  & 7353 & 1999 Dec 30.3--Jan 10.2 & \nodata & \nodata & 92
& \nodata \\ 
2000S  & 7384 & $<$Feb 28$^{b}$ & \nodata & \nodata & 94 & \nodata \\ 
2000cr & 7443 & Jun 21.2--25.9 & Jul 3.04 & 7.1--11.8 & 54 & No &
$\pm 42^{c}$ & \nodata & \nodata \\ 
2000de & 7478 & Jun 6--Aug 10.9 & \nodata & \nodata & 39 & \nodata \\ 
2000ds & 7507 & May 28--Oct 10.4 & \nodata & \nodata & 21 & \nodata \\ 
2000dt & 7508  & Sep 21--Oct 13.1 & \nodata & \nodata & 107 & \nodata \\
2000dv & 7510 & $<$Oct 17.1 & \nodata & \nodata & 63 & \nodata \\
2000ew & 7530 & May 8--Nov 28.5 & 2002 Jun 26 & 575--780 & 15 & No &
\nodata & \nodata & $\pm 67$ \\
2000fn & 7546 & Nov 1--19 & Dec 29.22 & 40.2--58.2 & 72 & No & $\pm
45$ & \nodata & \nodata \\ \\
2001B & 7555 & 2000 Dec 24.5--Jan 3.6 & Feb 4.49 &
31.9--42.0 & 24 & Yes & $3500\pm 29$ & \nodata & \nodata \\ 
      &      &                               & 2002 Oct 28.45 &
672.9--683 &    & No & $\pm 40$ & \nodata & \nodata \\  
2001M & 7568 & Jan 3.4--21.4 & Feb 4.45 & 14.1--32.1 & 56 & No & $\pm
28$ & \nodata & \nodata \\
%2001O(?)$^d$ & 7568 & $<$Jan 23.7 & Jan 30.25 & $\sim 22^e$ & 430 & No
%& $\pm 56$ & $\pm 70$ & $\pm 137$ \\
2001ai & 7605 & Mar 19.4--28.4 & Mar 31.36 & 3--12 & 118 & No &
$\pm 43^f$ & $\pm 49^f$ & $\pm 49^f$ \\
2001bb & 7614 & Apr 15.3--29.3 & May 5.26 & 6--20 & 72 & No &
$\pm 50$ & $\pm 60$ & $\pm 120$ \\
2001ch & 7637 & 2000 Nov 28.3--May 28.5 & \nodata & \nodata & 46
& \nodata \\
2001ci & 7618 & Apr 17.2--25.2 & Jun 6.35 & 42.2--50.2 & 17 & Yes &
$1450\pm 250$ & \nodata & \nodata \\
       &      &                & 2002 Jun 10.90 & 411.7--419.7 & & No
& $\pm 150$ & \nodata & \nodata \\
2001ef & 7710 & Aug 29--Sep 9.1 & Sep 14.3 & 5.2--16.3 & 38 & No &
$\pm 115^g$ & \nodata & \nodata \\
       &      &                 & 2002 Jun 18.0 & 281.9--293 & & No &
\nodata & \nodata & $\pm 88$ \\
2001ej & 7719 & Sep 1--17.1 & Sep 27.25 & 10.6--26.7 & 63 & No &
$\pm 44^h$ & \nodata & \nodata \\
       &      &                  & 2002 Jun 14.0+18.0 & 269.9--469 & & No &
\nodata & \nodata & $\pm 85$ \\
2001em & 7722  & Sep 10.3--15.3 & \nodata & \nodata & 91 & \nodata \\
2001eq(?) & 7728  & Aug 31.3--Sep 12.3 & \nodata & \nodata & 118 & \nodata \\ 
2001fw & 7751 & Oct 26--Nov 11.2 & \nodata & \nodata & 139 & \nodata \\
2001fx & 7751 & Oct 14.2--Nov 8.2 & \nodata & \nodata & 126 & \nodata \\
%2001gl & 7763 & Mar 29--Apr 15 & 2002 Jun 26.0 & 437--454 & 2070 & No &
%\nodata & \nodata & $\pm 90$ \\
2001is & 7782 & Dec 14--23 & 2002 Jun 14.0+18.0 & 175--184 & 60 & No
& \nodata & \nodata & $\pm 70^i$ \\ \\
2002J  & 7800 & Jan 9.4--21.4 & Jun 18.0 & 147.6--159.6 & 58 & No
& \nodata & \nodata & $\pm 85$ \\ 
2002ap$^j$ & 7810 & Jan 28.5 & Feb 1.03 & 3.5 & 7 & Yes &
$374\pm 29$ & \nodata & \nodata \\
2002bl & 7845 & Feb 14--Mar 2.9 & Mar 8.26 & 6--22 & 74 & No 
& $\pm 50$ & $\pm 45$ & \nodata \\ 
2002bm & 7845 & Jan 16--Mar 6.2 & Jun 26.0 & 111.8--161 & 85 & No &
\nodata & \nodata & $\pm 66$ \\ 
2002cg & 7877 & Mar 28.5--Apr 13.5 & Jun 26.0 & 73.5--89.5 & 150 & No &
\nodata & \nodata & $\pm 74^k$ \\ 
2002cj & 7882 & Apr 9.5--21.5 & Jun 3.19 & 42.7--54.7 & 106 & Yes &
$112\pm 33$ & $220\pm 37$ & $240\pm 48$ \\
       &      &               & Jun 18.0 & 57.5--69.5 &     & Yes & 
\nodata & \nodata & $408\pm 81$ \\ 
       &      &               & Jun 26.0 & 65.5--77.5 &     & Yes & 
\nodata & \nodata & $300\pm 68$ \\ 
2002cp & 7887 & Apr 11.2--28.2 & Jun 1.96 & 34.8--51.8 & 80 & No &
$\pm 31$ & $\pm 35$ & $\pm 36$ \\ 
2002cw & 7902 & 2001 Oct 1.2--May 16.5 & \nodata & \nodata & 71 &
\nodata \\ 
2002dg & 7915 & 2002 May 5--May 31.3 & Jul 9.95 & 39.7--66 & 215$^l$
& Hint & $\pm 33$ & $92\pm 37$ & $\pm 46$ \\
2002dn(?) & 7922 & May 31.5--Jun 15.5 & Jul 4.4 & 18.9--33.9 & 115 &
No & $\pm 39$ & $\pm 43$ & $\pm 72$ \\
2002dz & 7935 & 2001 Nov 10.2--Jul 16.5 & \nodata & \nodata &
84 & \nodata \\
2002ex & 7964 & Aug 19.3 & \nodata & \nodata & 180 & \nodata \\
2002fh(?) & 7971 & Apr 9--May 9.3 & \nodata & \nodata & 1870 & \nodata \\ 
2002ge(?)$^m$ & 7987 & 2000 Oct 1.8--2002 Oct 7.9 & Oct 12.0 & $\sim
13$ & 47 & No & $\pm 160^n$ & $\pm 130^o$ & \nodata \\
2002gy    & 7996 & Oct 9.4--16.4 & Oct 28.45 & 19.1--26.1 & 114 & No &
$\pm 36$ & $\pm39$ & \nodata \\
2002hf    & 8004 & Oct 22.3--29.3 & Nov 7.17 & 8.9--16.9 & 88 & No &
$\pm 40$ & \nodata & \nodata \\
2002hn    & 8009 & Oct 21.5--30.5 & Nov 7.42 & 7.9--16.9 & 82 & No & 
$\pm 44$ & \nodata & \nodata \\
2002ho    & 8011 & May 27--Nov 5.1 & Nov 15.33 & $\sim 56^p$ & 
42 & No & $\pm 47$ & $\pm 54$ & \nodata  \\
2002hy    & 8016 & Oct 13.1--Nov 12.1 & Nov 21.71 & 9.6--39.6 & 58 &
No & $\pm 74$ & \nodata & \nodata \\
2002hz    & 8017 & Nov 2.2--12.2        & 2003 Jan 21.09 & 69.9--70.9
& 85 & No & $\pm 29$ & \nodata & \nodata \\
2002ji	  & 8025 & Apr 10--Nov 30.8 & Dec 5.67 & 38.4--59.4$^q$ & 23 &
No & $\pm 43$ & \nodata & \nodata \\
2002jj	  & 8026 & Oct 1.4--24.4 & Dec 15.28 & 51.9--74.9 & 66 &
No & $\pm 33$ & \nodata & \nodata \\
2002jp	  & 8031 & May 14.2--Nov 23.5 & Dec 14.55 & $\sim 33^r$ &
58 & No & $\pm 38$ & \nodata & \nodata \\
2002jz    & 8037 & 2001 Dec 5--2002 Dec 23.3 & 2003 Jan 3.28 & $\sim
32^s$ & 24 & No & $\pm 25^t$ & \nodata & \nodata \\  
\enddata
\tablecomments{The columns are (left to right): (1) SN name, (2) IAU
Circular number for the initial detection, (3) time of the SN
explosion, with the range given between the most recent observation of
the galaxy which did not show the SN and the epoch at which the SN was
detected, (4) epoch of our VLA observations, (5) time delay between the
SN explosion and the epoch of our observations, (6) distance to the
galaxy (assuming $H_0=65$ km s$^{-1}$ Mpc$^{-1}$), (7) indicates
whether radio emission was detected, (8) flux density at 8.46 GHz, (9)
flux density at 4.86 GHz, and (10) flux density at 1.43 GHz. \newline
$^a$ uncertainties are quoted as $1\sigma$ rms; $^b$ nebular phase;
$^c$ falls on top of galaxy substructure, $<10\sigma$; $^d$ Ia or Ic
(IAUC 7574); $^e$ before maximum (IAUC 7574); $^f$ falls on top of
galaxy with complex substructure, $<4\sigma$;  $^g$ falls on
top of galaxy, $<5\sigma$; $^h$ falls on top of galaxy, $<35\sigma$;
$^i$ on top of galaxy substructure, $<3\sigma$; $^j$ See
\citet{bkc02}; $^k$ on top of galaxy, $<10\sigma$;  $^l$ IAUC 7922;
$^m$ Type Ic similar to SN\,1994I a few days before maximum, or a
sub-luminous Type Ia (IAUC 7990); $^n$ on top of galaxy, $<10\sigma$;
$^o$ on top of galaxy, $<3\sigma$; $^p$ within a few weeks past
maximum brightness (IAUC 8014); $^q$ $3-6$ weeks past maximum light
(IAUC 8026); $^r$ two weeks past maximum light (IAUC 8031); $^s$ ten
days past maximum (IAUC 8037); $^t$ on top of galaxy substructure,
$<3\sigma$.}      
\end{deluxetable}

\begin{deluxetable}{lcccccc}
\tablecolumns{7}
\tablewidth{0pc}
\tablecaption{Ejecta and Progenitor Properties of Type Ib/c Supernovae
Detected in the Radio\label{tab:ejecta}} 
\tablehead{
\colhead {SN} &
\colhead {$E_K$} &
\colhead {$M_{\rm ej}$} &
\colhead {$M_{\rm ^{56}Ni}$} &
\colhead {$M_{\rm CO}$} &
\colhead {$M_{\rm prog}$} &
\colhead {Ref.} \\
\colhead {} &
\colhead {($10^{51}$ erg)} &
\colhead {(M$_\odot$)} &
\colhead {(M$_\odot$)} &
\colhead {(M$_\odot$)} &
\colhead {(M$_\odot$)} &
\colhead {}
}
\startdata
%1983N  & \nodata & \nodata & \nodata & \nodata & \nodata & \nodata \\
1984L  & $20$    & $50$    & $0.2$   & \nodata & \nodata & 1 \\
       & \nodata & $10$    & \nodata & \nodata & \nodata & 2 \\
       & \nodata & \nodata & $0.1$   & \nodata & $20-30$ & 3 \\
%1990B  & \nodata & \nodata & \nodata & \nodata & \nodata & \nodata \\
1994I  & $1-1.4$ & $0.9-1.3$ & 0.07  & $<1.5$  & $<15$   & 4 \\
       & \nodata & $0.9$     & 0.07  & $1.35$  & \nodata & 5 \\
       & 1       & \nodata & \nodata & \nodata & \nodata & 6 \\
1998bw & 30      & \nodata & 0.7     & 13.8    & 40      & 7 \\
       & 50      & 10      & 0.4     & 13.8    & 40      & 8 \\
       & 22      & \nodata & 0.5     & 6.5     & \nodata & 9 \\
       & 2       & 2       & 0.2     & \nodata & \nodata & 10 \\
       & \nodata & \nodata & $0.5-0.9^a$ & \nodata & \nodata & 11 \\
2002ap & $4-10$  & $2.5-5$ & 0.07    & 5       & $20-25$ & 12 
\enddata
\tablecomments{The columns are (left to right): (1) Ejecta kinetic
energy, (2) ejecta mass, (3) $^{56}$Ni mass, (4) estimated mass of the
CO core, (5) estimated mass of the progenitor, and (6) references.
Data are not available for the SNe detected in this survey.  $^a$
These authors use the models of \citet{imn+98} and \citet{wes99} as
input; they assert that observations in the nebular phase exclude the
low $^{56}$Ni mass inferred by \citet{hww99}.}
\tablerefs{(1) \citet{byb93}; (2) \citet{sw91}; (3) \citet{sk89}; (4)
\citet{ybb95}; (5) \citet{inh+94}; (6) \citet{mbb+99}; (7)
\citet{imn+98}; (8) \citet{nmn+01}; (9) \citet{wes99}; (10)
\citet{hww99}; (11) \citet{skf+00}; (12) \citet{mdm+02}}
\end{deluxetable}

\begin{deluxetable}{llcc}
\tablecolumns{4}
\tablewidth{0pc}
\tablecaption{Best-Fit Models for the Supernova and $\gamma$-Ray Burst
Populations\label{tab:bayes}}
\tablehead{
\colhead {Population} &
\colhead {Model}      &
\colhead {Parameters} &
\colhead {${\rm log}({\cal L})/{\rm dof}$}
}
\startdata
SN	& Gaussian	& (26.1, 1.0)		& 22.5/36 \\
SN	& D.~Powerlaw	& ($-29.0$, 25.4)	& 22.4/36 \\
SN	& Flat		& (20.0, 29.1)		& 22.3/36 \\
GRB	& Gaussian	& (31.0, 0.8) 		& 38.8/33 \\
GRB	& D.~Powerlaw	& ($-22.5$, 29.6) 	& 48.9/33 \\
GRB	& Flat		& (29.6, 32.4)		& 37.7/33 \\
SN+GRB	& Gaussian	& (28.4, 2.4)   	& 130.3/71 \\
SN+GRB	& D.~Powerlaw	& ($-10.4$, 25.3)  	& 123.8/71 \\
SN+GRB	& I.~Powerlaw	& (3.7, 32.3)		& 103.7/71 \\
SN+GRB	& Flat		& (22.6, 32.4)  	& 104.9/71 \\
\enddata
\tablecomments{The columns are (left to right): (1) Data set, (2)
population distribution function, (3) best-fit parameters, and (4) log
likelihood.  A detailed explanation of the models and the fitting
procedure is provided in \S\ref{sec:comp}.}
\end{deluxetable}

\clearpage
\begin{figure} 
\plotone{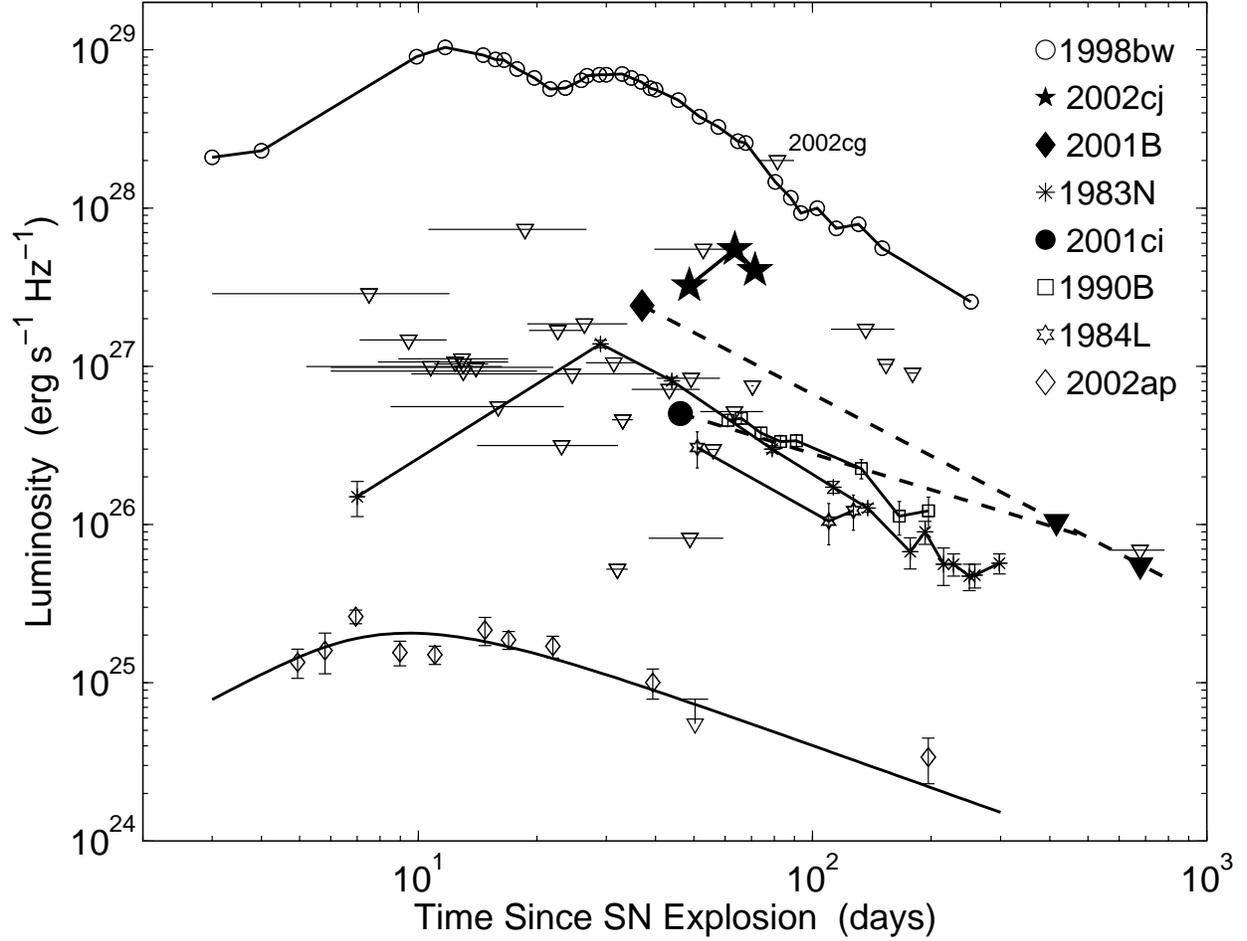}
\caption{Radio lightcurves of Type Ib/c SNe detected in this survey
and from the literature, as well as upper limits for the
non-detections; these are plotted as $3\sigma$ in most cases with the
exception of SNe which are located on top of a bright host galaxy (see
Table~\ref{tab:data}).  For SN\,2002ap we plot the model of
\citet{bkc02}, while the other solid lines simply trace the
observations and do not represent a model fit.  The uncertainty in
time for the non-detections represents the uncertain time of 
explosion.  We note that for SN\,2002cg, which is the only SN that is
potentially brighter than SN\,1998bw, the limit is $10\sigma$ due to
the superposition of the SN on top of its host galaxy.  
\label{fig:lcs}}
\end{figure}

\clearpage
\begin{figure} 
\epsscale{0.8}
\plotone{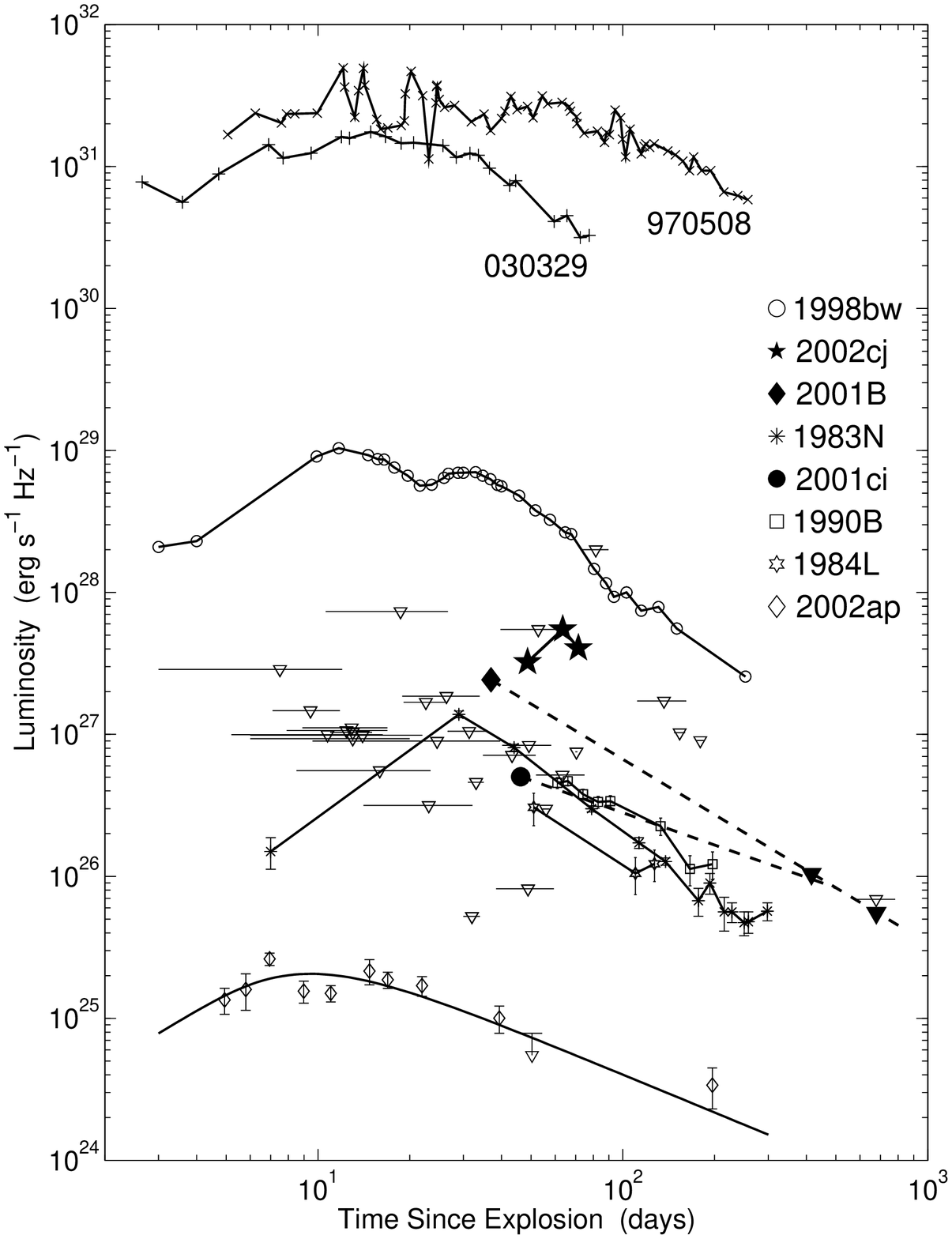}
\caption{Same as Figure~\ref{fig:lcs} but including the 8.46 GHz 
lightcurves of GRB\,970508 \citep{fwk00} and GRB\,030329
\citep{bkp+03}.  These GRB afterglows are at least two orders of
magnitude brighter than SN\,1998bw, the brightest Type Ib/c SN.  We
note that the fluctuations in the GRB lightcurves are not intrinsic,
and arise instead from interstellar scintillation.  Based on the
significant difference in radio luminosity we rule out the possibility
that the Type Ib/c SN observed here produced a GRB.  This is discussed
more quantitatively in in \S\ref{sec:comp} and Figures~\ref{fig:hist1}
and \ref{fig:hist2}.
\label{fig:lcs2}}
\end{figure}

\clearpage
\begin{figure}
\epsscale{1}
\plotone{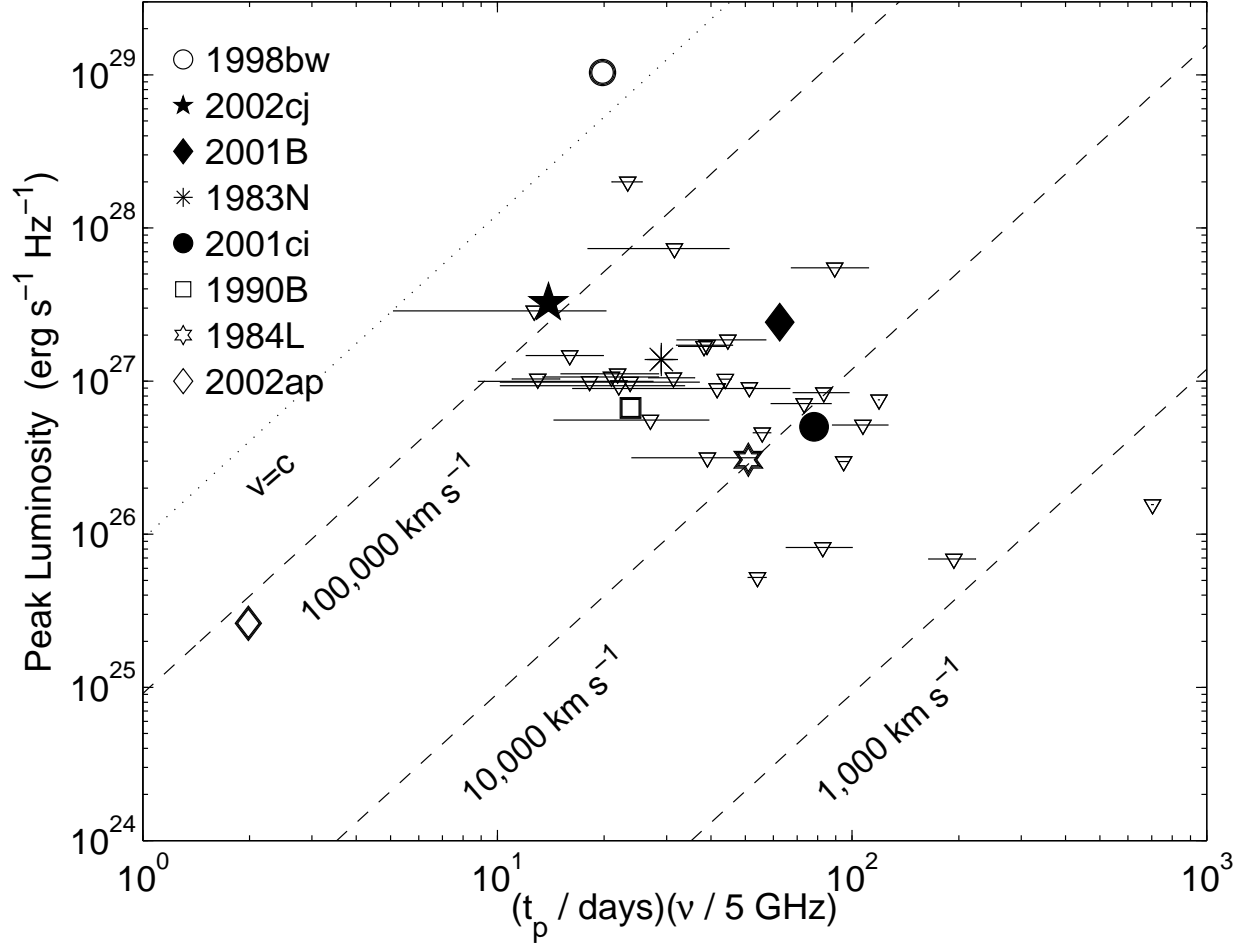}
\caption{Peak radio luminosity plotted versus the time of peak
luminosity for Ib/c SNe studied in this survey and from the
literature.  Symbols are as in Figure~\ref{fig:lcs}.  The diagonal
lines are contours of constant average expansion velocity based on the
assumption that the peak of the radio luminosity occurs at the
synchrotron self-absorption frequency \citep{che98}.  While the upper
limits do not necessarily measure the peak of the spectrum at the time
of the observation, a comparison to SN\,1983N indicates that the range
of time delays relative to the SN explosion reasonably samples the
peak.  Upper limits measured at $t\gtrsim 100$ days probably miss the
peak of the synchrotron spectrum and therefore do not provide a useful
limit. 
\label{fig:vel}}
\end{figure}

\clearpage

\clearpage

\clearpage


\begin{thebibliography}{}

\bibitem[{Baron}, {Young} \& {Branch}(1993)]{byb93}
{Baron}, E., {Young}, T.~R., and {Branch}, D. 1993, \apj, 409, 417.

\bibitem[{Berger}, {Kulkarni} \& {Chevalier}(2002)]{bkc02}
{Berger}, E., {Kulkarni}, S.~R., and {Chevalier}, R.~A. 2002, \apjl, 577, L5.

\bibitem[{Berger}, {Kulkarni} \& {Frail}(2003)]{bkf03}
{Berger}, E., {Kulkarni}, S.~R., and {Frail}, D.~A. 2003, ApJ, 590, 379.

\bibitem[{Berger} {\it et al.}\ (2003)]{bkp+03}
{Berger}, E. {\it et al.}\  2003, {in prep.}

\bibitem[{Bloom}, Frail \& Kulkarni(2003)]{bfk03}
{Bloom}, J.~S., Frail, D.~A., and Kulkarni, S.~R. 2003, astro-ph/0302210.

\bibitem[{Bloom} {\it et al.}\ (1998)]{bkh+98}
{Bloom}, J.~S., {Kulkarni}, S.~R., {Harrison}, F., {Prince}, T., {Phinney},
  E.~S., and {Frail}, D.~A. 1998, \apjl, 506, L105.

\bibitem[{Branch}, {Nomoto} \& {Filippenko}(1991)]{bnf91}
{Branch}, D., {Nomoto}, K., and {Filippenko}, A.~V. 1991, Comm.~Astrophys., 15,
  221.

\bibitem[{Cappellaro}, {Evans} \& {Turatto}(1999)]{cet99}
{Cappellaro}, E., {Evans}, R., and {Turatto}, M. 1999, \aap, 351, 459.

\bibitem[{Chevalier}(1982)]{che82}
{Chevalier}, R.~A. 1982, \apj, 258, 790.

\bibitem[{Chevalier}(1998)]{che98}
{Chevalier}, R.~A. 1998, \apj, 499, 810.

\bibitem[{Chornock}(2002)]{cho02-iauc7894}
{Chornock}, R. 2002, \iaucirc, 7894, 2.

\bibitem[{Chornock} \& {Filippenko}(2001)]{cf01-iauc7577}
{Chornock}, R. and {Filippenko}, A.~V. 2001, \iaucirc, 7577, 2.

\bibitem[{Clocchiatti} {\it et al.}\ (2000)]{cps+00}
{Clocchiatti}, A. {\it et al.}\  2000, \apj, 529, 661.

\bibitem[{Colgate}(1968)]{col68}
{Colgate}, S.~A. 1968, Canadian Journal of Physics, 46, 476.

\bibitem[{Filippenko}(1997)]{fil97}
{Filippenko}, A.~V. 1997, \araa, 35, 309.

\bibitem[{Filippenko} \& {Chornock}(2001)]{fc01-iauc7638}
{Filippenko}, A.~V. and {Chornock}, R. 2001, \iaucirc, 7638, 1.

\bibitem[{Folkes} {\it et al.}\ (1999)]{frp+99}
{Folkes}, S. {\it et al.}\  1999, \mnras, 308, 459.

\bibitem[{Fomalont}(1981)]{fom81}
{Fomalont}, E. 1981, NEWSLETTER.~NRAO NO.~3, P.~3, 1981, 3, 3.

\bibitem[{Frail} {\it et al.}\ (2003)]{fkb+03}
{Frail}, D.~A., {Kulkarni}, S.~R., {Berger}, E., and {Wieringa}, M.~H. 2003,
  \aj, 125, 2299.

\bibitem[{Frail} {\it et al.}\ (2001)]{fks+01}
{Frail}, D.~A. {\it et al.}\  2001, \apjl, 562, L55.

\bibitem[{Frail}, {Waxman} \& {Kulkarni}(2000)]{fwk00}
{Frail}, D.~A., {Waxman}, E., and {Kulkarni}, S.~R. 2000, \apj, 537, 191.

\bibitem[{Galama} {\it et al.}\ (1998)]{gvv+98}
{Galama}, T.~J. {\it et al.}\  1998, \nat, 395, 670.

\bibitem[{Ganeshalingam} \& {Li}(2002)]{gl02-iauc7882}
{Ganeshalingam}, M. and {Li}, W.~D. 2002, \iaucirc, 7882, 1.

\bibitem[{Garnavich} {\it et al.}\ (1998)]{gjk+98}
{Garnavich}, P., {Jha}, S., {Kirshner}, R., and {Berlind}, P. 1998, \iaucirc,
  7066, 1.

\bibitem[{Germany} {\it et al.}\ (2000)]{grs+00}
{Germany}, L.~M., {Reiss}, D.~J., {Sadler}, E.~M., {Schmidt}, B.~P., and
  {Stubbs}, C.~W. 2000, \apj, 533, 320.

\bibitem[{Goodman}(1986)]{goo86}
{Goodman}, J. 1986, \apjl, 308, L47.

\bibitem[{Granot} \& Loeb(2003)]{gl03}
{Granot}, J. and Loeb, A. 2003, astro-ph/0305379.

\bibitem[{Granot} {\it et al.}\ (2002)]{gpk+02}
{Granot}, J., {Panaitescu}, A., {Kumar}, P., and {Woosley}, S.~E. 2002, \apjl,
  570, L61.

\bibitem[{H{\" o}flich}, {Wheeler} \& {Wang}(1999)]{hww99}
{H{\" o}flich}, P., {Wheeler}, J.~C., and {Wang}, L. 1999, \apj, 521, 179.

\bibitem[{Hamuy}(2003)]{ham03}
{Hamuy}, M. 2003, \apj, 582, 905.

\bibitem[{Heise}, {in 't Zand} \& {Kulkarni}(2003)]{hzk03}
{Heise}, J., {in 't Zand}, J.~J.~M., and {Kulkarni}, S.~R. 2003, {in prep.}

\bibitem[{Hjorth} \& et~al.(2003)]{hjo+03}
{Hjorth}, J. and et~al. 2003, \nat, 423, 847.

\bibitem[{Iwamoto} {\it et al.}\ (1998)]{imn+98}
{Iwamoto}, K. {\it et al.}\  1998, \nat, 395, 672.

\bibitem[{Iwamoto} {\it et al.}\ (2000)]{inn+00}
{Iwamoto}, K. {\it et al.}\  2000, \apj, 534, 660.

\bibitem[{Iwamoto} {\it et al.}\ (1994)]{inh+94}
{Iwamoto}, K., {Nomoto}, K., {Hoflich}, P., {Yamaoka}, H., {Kumagai}, S., and
  {Shigeyama}, T. 1994, \apjl, 437, L115.

\bibitem[{Kawabata} {\it et al.}\ (2002)]{kji+02}
{Kawabata}, K.~S. {\it et al.}\  2002, \apjl, 580, L39.

\bibitem[{Kulkarni} {\it et al.}\ (1998)]{kfw+98}
{Kulkarni}, S.~R. {\it et al.}\  1998, \nat, 395, 663.

\bibitem[{Li} \& {Chevalier}(1999)]{lc99}
{Li}, Z. and {Chevalier}, R.~A. 1999, \apj, 526, 716.

\bibitem[{MacFadyen} \& {Woosley}(1999)]{mw99}
{MacFadyen}, A.~I. and {Woosley}, S.~E. 1999, \apj, 524, 262.

\bibitem[{Marzke} {\it et al.}\ (1998)]{mcp+98}
{Marzke}, R.~O., {da Costa}, L.~N., {Pellegrini}, P.~S., {Willmer}, C.~N.~A.,
  and {Geller}, M.~J. 1998, \apj, 503, 617.

\bibitem[{Matheson} {\it et al.}\ (2001)a]{mfl+01}
{Matheson}, T., {Filippenko}, A.~V., {Li}, W., {Leonard}, D.~C., and {Shields},
  J.~C. 2001a, \aj, 121, 1648.

\bibitem[{Matheson} {\it et al.}\ (2001)b]{mjc+01-iauc7563}
{Matheson}, T., {Jha}, S., {Challis}, P., {Kirshner}, R., and {Calkins}, M.
  2001b, \iaucirc, 7563, 2.

\bibitem[{Matheson} {\it et al.}\ (2002)]{mjc+02-iauc7894}
{Matheson}, T., {Jha}, S., {Challis}, P., {Kirshner}, R., {Calkins}, M.,
  {Chornock}, R., {Li}, W.~D., and {Filippenko}, A.~V. 2002, \iaucirc, 7894, 1.

\bibitem[{Matzner} \& {McKee}(1999)]{mm99}
{Matzner}, C.~D. and {McKee}, C.~F. 1999, \apj, 510, 379.

\bibitem[{Mazzali} {\it et al.}\ (2002)]{mdm+02}
{Mazzali}, P.~A. {\it et al.}\  2002, \apjl, 572, L61.

\bibitem[{Mazzali}, {Iwamoto} \& {Nomoto}(2000)]{min00}
{Mazzali}, P.~A., {Iwamoto}, K., and {Nomoto}, K. 2000, \apj, 545, 407.

\bibitem[{Millard} {\it et al.}\ (1999)]{mbb+99}
{Millard}, J. {\it et al.}\  1999, \apj, 527, 746.

\bibitem[{Nakamura}(1999)]{nak99}
{Nakamura}, T. 1999, \apjl, 522, L101.

\bibitem[{Nakamura} {\it et al.}\ (2001)]{nmn+01}
{Nakamura}, T., {Mazzali}, P.~A., {Nomoto}, K., and {Iwamoto}, K. 2001, \apj,
  550, 991.

\bibitem[{Nomoto} {\it et al.}\ (1994)]{nyp94}
{Nomoto}, K., {Yamaoka}, H., {Pols}, O.~R., {van den Heuvel}, E.~P.~J.,
  {Iwamoto}, K., {Kumagai}, S., and {Shigeyama}, T. 1994, \nat, 371, 227.

\bibitem[{Norris}(2002)]{nor02}
{Norris}, J.~P. 2002, \apj, 579, 386.

\bibitem[{Paczynski}(1986)]{pac86}
{Paczynski}, B. 1986, \apjl, 308, L43.

\bibitem[{Paczynski}(1998)]{pac98}
{Paczynski}, B. 1998, \apjl, 494, L45+.

\bibitem[{Paczynski}(2001)]{pac01}
{Paczynski}, B. 2001, Acta Astronomica, 51, 1.

\bibitem[{Panagia}, {Sramek} \& {Weiler}(1986)]{psw+86}
{Panagia}, N., {Sramek}, R.~A., and {Weiler}, K.~W. 1986, \apjl, 300, L55.

\bibitem[{Panaitescu} \& {Kumar}(2002)]{pk02}
{Panaitescu}, A. and {Kumar}, P. 2002, \apj, 571, 779.

\bibitem[{Pian} {\it et al.}\ (2000)]{paa+00}
{Pian}, E. {\it et al.}\  2000, \apj, 536, 778.

\bibitem[{Readhead}(1994)]{rea94}
{Readhead}, A.~C.~S. 1994, \apj, 426, 51.

\bibitem[{Reichart} \& Yost(2001)]{ry01}
{Reichart}, D.~E. and Yost, S. 2001, astro-ph/0107545.

\bibitem[{Rigon} {\it et al.}\ (2003)]{rtb+03}
{Rigon}, L. {\it et al.}\  2003, \mnras, 340, 191.

\bibitem[{Rossi}, {Lazzati} \& {Rees}(2002)]{rlr02}
{Rossi}, E., {Lazzati}, D., and {Rees}, M.~J. 2002, \mnras, 332, 945.

\bibitem[{Ruderman}, {Tao} \& {Klu{\' z}niak}(2000)]{rtk00}
{Ruderman}, M.~A., {Tao}, L., and {Klu{\' z}niak}, W. 2000, \apj, 542, 243.

\bibitem[{Schlegel} \& {Kirshner}(1989)]{sk89}
{Schlegel}, E.~M. and {Kirshner}, R.~P. 1989, \aj, 98, 577.

\bibitem[{Schmidt}(2001)]{sch01}
{Schmidt}, M. 2001, \apj, 552, 36.

\bibitem[{Sollerman} {\it et al.}\ (2000)]{skf+00}
{Sollerman}, J., {Kozma}, C., {Fransson}, C., {Leibundgut}, B., {Lundqvist},
  P., {Ryde}, F., and {Woudt}, P. 2000, \apjl, 537, L127.

\bibitem[{Sramek}, {Panagia} \& {Weiler}(1984)]{spw84}
{Sramek}, R.~A., {Panagia}, N., and {Weiler}, K.~W. 1984, \apjl, 285, L59.

\bibitem[{Stanek} {\it et al.}\ (2003)]{smg+03}
{Stanek}, K.~Z. {\it et al.}\  2003, ArXiv Astrophysics e-prints, 4173.

\bibitem[{Swartz} \& {Wheeler}(1991)]{sw91}
{Swartz}, D.~A. and {Wheeler}, J.~C. 1991, \apjl, 379, L13.

\bibitem[{Swift}, {Li} \& {Filippenko}(2001)]{slf01-iauc7618}
{Swift}, B., {Li}, W.~D., and {Filippenko}, A.~V. 2001, \iaucirc, 7618, 1.

\bibitem[Totani(2003)]{tot03}
Totani, T. 2003, astro-ph/0303621.

\bibitem[{Totani} \& {Panaitescu}(2002)]{tp02}
{Totani}, T. and {Panaitescu}, A. 2002, \apj, 576, 120.

\bibitem[{Uomoto}(1986)]{uom86}
{Uomoto}, A. 1986, \apjl, 310, L35.

\bibitem[{van Dyk} {\it et al.}\ (1993)]{dsw+93}
{van Dyk}, S.~D., {Sramek}, R.~A., {Weiler}, K.~W., and {Panagia}, N. 1993,
  \apj, 409, 162.

\bibitem[{Weiler} {\it et al.}\ (1986)]{wsp+86}
{Weiler}, K.~W., {Sramek}, R.~A., {Panagia}, N., {van der Hulst}, J.~M., and
  {Salvati}, M. 1986, \apj, 301, 790.

\bibitem[{Woosley}(1993)]{woo93}
{Woosley}, S.~E. 1993, \apj, 405, 273.

\bibitem[{Woosley}, {Eastman} \& {Schmidt}(1999)]{wes99}
{Woosley}, S.~E., {Eastman}, R.~G., and {Schmidt}, B.~P. 1999, \apj, 516, 788.

\bibitem[{Woosley}, {Langer} \& {Weaver}(1993)]{wlw93}
{Woosley}, S.~E., {Langer}, N., and {Weaver}, T.~A. 1993, \apj, 411, 823.

\bibitem[{Woosley} \& {Weaver}(1986)]{ww86}
{Woosley}, S.~E. and {Weaver}, T.~A. 1986, \araa, 24, 205.

\bibitem[{Xu} \& {Qiu}(2001)]{xq01-iauc7555}
{Xu}, D.~W. and {Qiu}, Y.~L. 2001, \iaucirc, 7555, 2.

\bibitem[{Young}, {Baron} \& {Branch}(1995)]{ybb95}
{Young}, T.~R., {Baron}, E., and {Branch}, D. 1995, \apjl, 449, L51+.

\bibitem[{Zhang} \& {M{\' e}sz{\' a}ros}(2002)]{zm02}
{Zhang}, B. and {M{\' e}sz{\' a}ros}, P. 2002, \apj, 571, 876.

\end{thebibliography}
\end{document}